\documentclass[english,american,aps,prl,twocolumn,showpacs]{revtex4}
\usepackage[T1]{fontenc}
\usepackage[latin1]{inputenc}
\usepackage{graphicx}
\makeatletter
\usepackage[scanall]{psfrag}
\usepackage{babel}
\makeatother

\begin{document}

\title{Microcanonical versus Canonical Analysis of Protein Folding}

\author{J.~Hern\'andez-Rojas}

\email{jhrojas@ull.es}

\author{J.~M.~Gomez~Llorente}

\email{jmgomez@ull.es}
 
\affiliation{Departamento de F\'{\i}sica Fundamental II and IUDEA. 
Universidad de La Laguna.
38205 Tenerife. Spain }

\begin{abstract}
The microcanonical analysis is shown to be a powerful tool to characterize
the protein folding transition and to neatly distinguish between good
and bad folders. An off-lattice model with parameter chosen to represent
polymers of these two types is used to illustrate this approach. Both
canonical and microcanonical ensembles are employed. The required
calculations were performed using parallel tempering Monte Carlo simulations.
The most revealing features of the folding transition are related
to its first-order-like character, namely, the S-bend pattern in the
caloric curve, which gives rise to negative microcanonical specific
heats, and the bimodality of the energy distribution function at the
transition temperatures. Models for a good folder are shown to be
quite robust against perturbations in the interaction potential parameters.
\end{abstract}


\maketitle
A protein is a polypeptide, i.e., a sequence of amino acids. Only a
very small fraction of all possible sequences are valid,
that is, they determine uniquely the native state. This state is the
three-dimensional compact conformation that is stable and functionally
active at room temperature in its biological environment \cite{Anfinsen73}.
Protein folding, i.e., the physical process by which a polypeptide
folds from low-structure denatured states into its native state, remains
an open fundamental problem in molecular biology. This process is
remarkably fast ($10^{-3}$ to $1$ sec.) for natural proteins, which
are the good folders selected by biological evolution. The understanding
of the structural and energetic properties that characterize a good
folder is a basic issue in the protein folding problem an a necessary
step towards two ambitious goals, namely, the prediction of the native
structure from the amino acid sequence, and the corresponding inverse
design problem. The formulation of energy landscape theory
was a major breakthrough in this research field \cite{Wolynes87}.
In this approach, good folders such as the natural proteins
present globally funneled potential energy surfaces;
this implies that the native state
can be directly reached from almost any initial (unfolded) structure.
The folding funnel landscape is closely related to the principle of minimal frustration
\cite{Wolynes87}. High frustration is a feature characterizing the energy landscape
of homopolymers, which are, therefore, generally
bad folders.

Proteins having a well defined and unique
native state which is reached from almost any initial condition
induce a thermodynamic behavior which is similar to that of small
systems undergoing phase separation between the denatured macrostate and the
native one. This two-phase picture is also essential in the kinetic
analysis of folding, where each of the two phases is associated with
a free energy minimum along the reaction coordinate, and the transition
is described as an activated rate process. The very different energy
of the unfolded and native states is the origin of the two symmetry
unrelated thermodynamic phases; thus the folding transition should
be of first order in the thermodynamic limit. The two-phase picture
is however inadequate to the corresponding transition in bad folders
such as homopolymers, which is known to be a second order transition
(the so called coil-globule or $\theta$-transition) \cite{Dill91}.
The order of the transition is therefore a determining factor to characterize
good and bad folders in the thermodynamical limit. However, proteins
are actually finite size systems, and the classification of the transition
under these circumstances is not an obvious task. Schemes based on
Fisher zeroes or Lee-Yang zeros \cite{Fisher65,Lee52} of the partition
function depending on a complex intensive parameter (e.g. temperature)
have been used to analyze the thermodynamics of finite-size systems
\cite{Harting00}, and more recently these methods have been applied
to the characterization of the protein folding transition \cite{Wang03}.

Closely connected to the behavior of Lee-Yang zeros in the complex
temperature plane, some features of the microcanonical ensemble
have turned out to be very useful in the characterization
of transitions in finite-size
systems where phase separation appear as a relevant pattern, as seem
to occur in the folding transition for good folders. These systems
may become colder upon absorbing energy (S-bend), which would lead
to a convex behavior of the microcanonical entropy and to negative
values of the corresponding specific heat. These features are rigorously linked
to the bimodality of the energy distribution function at the transition
temperature \cite{Wales94}, again a clear evidence of the two-phase coexistence.
These characteristic phenomena have been observed in astrophysical
systems \cite{Thirring70}, in
magic-number clusters \cite{Wales94}, in nuclei fragmentation \cite{Pichon05},
and in finite-lattice spin models \cite{Pleimling06}.
Related to the present work,
the microcanonical analysis
of peptide aggregation processes \cite{Janke06} and of association
of hydrophobic segments in heteropolymers \cite{Liang07} have been carried
out more recently.

In this Letter, we will demonstrate that the microcanonical analysis
is able to characterize in a precise way the protein folding transition,
thus allowing us to neatly distinguish between good and bad folders.
Therefore, this scheme would provide us with a new tool in protein
folding research that is simpler than the schemes based on the analysis
of either the energy landscape or the behavior of partition function
zeros in the complex temperature plane, but as powerful as them.

In order to illustrate how to characterize protein folding by the
microcanonical analysis, the simple but still realistic off-lattice
model proposed by Clementi \textit{et al}. \cite{Clementi98} (with some slight
modifications) will be employed. This gives the conformation of a
sequence of $N$ beads (each one representing a residue) by their
coordinates $\mathbf{r}_{1},\ldots\mathbf{r}_{N}$ in three-dimensional
space. The potential energy surface (PES) of the polymer is built
as a sum of pairwise interactions $U_{ij}=\delta_{i,j+1}a(r_{ij}-r_{0})^{2}+(1-\delta_{i,j+1})4\epsilon_{ij}[(\sigma_{ij}/r_{ij})^{12}-(\sigma_{ij}/r_{ij})^{6}]$,
where the interaction between two non-bonding beads is chosen as a
Lennard-Jones potential, while between bonding beads we have used
a harmonic potential with force constant $a=50$ $\textrm{\AA}^{-2}$
(energies shall be measured in arbitrary dimensionless units) and
equilibrium distance $r_{0}=3.8$ $\textrm{\AA}$. In our analysis
we shall consider the two polymer models with $N=30$ proposed earlier
\cite{Clementi98}, and a third new model.
The first one corresponds to a homopolymer (HMP) for
which we have chosen, $\epsilon_{ij}=10$ and $\sigma_{ij}=6.5$ $\textrm{\AA}$
for all $i,\: j$. As expected for this system, this set of parameters
leads to a very rugged energy landscape and therefore to a high frustration
system; all these features characterize a bad folder. With the second
model, a designed heteropolymer (DHTP), we intend to represent a natural protein, i.e. a good folder.
In this case the proposed polymer sequence \cite{Clementi98} is
intended to make coincide the protein native structure with the homopolymer
global minimum. As far as the $\epsilon_{ij}$ Lennard-Jones parameter
are concerned, this sequence is made with four types of amino acids
and, therefore, 10 different $\epsilon_{ij}$ values in the range $0.25\leq\epsilon_{ij}\leq10$.
Following our reference work \cite{Clementi98}, we have chosen the
$\sigma_{ij}$ values so that the native structure corresponding to
the chosen sequence is significantly stabilized, i.e. its frustration
minimized; we obtain in this way values in the range $5<\sigma_{ij}<17$
$\textrm{\AA}$. This parameter set leads to a clear funneled landscape,
which characterizes good folders. The third model has also $N=30$ and it
is a random heteropolymer (RHTP)
given by a random sequence of the previous
four amino acids with the same set of $\epsilon_{ij}$ values and $\sigma_{ij}=6.5$ $\textrm{\AA}$.

At this point we should mention
that homopolymers models different from
those chosen in this work have
been either shown or suggested to present a first-order-like (liquid-solid-like)
transition at a temperature below the $\theta$-like transition \cite{Zhou96,Maritan04}.
Parallelly, protein models have been proposed that present a $\theta$-like
transition above the folding transition temperature \cite{Dill95}.
In these cases, the two $\theta$-transitions as well as the protein
folding and homopolymer liquid-to-solid transitions can be identified,
but the major difference between the latter two is that the folded solid-like
states in the homopolymer present polymorphism, while the protein
native state is generally unique. Therefore, the mechanisms behind these two
first-order-like transitions are different. Under these circumstances our thermodynamic analysis will
not be able to neatly distinguish between good and bad folders, but
checking for the uniqueness of the native state would solve the issue.
Indeed, using the basin-hopping method\cite{Wales97,Li87} we have checked
that polymorphism also characterizes the model
introduced to study the association of
hydrophobic segments in heteropolymers \cite{Liang07}.
Therefore, one can not draw any conclusions about the
folding transition from that study.

We have performed parallel tempering (also known as replica exchange)
Monte Carlo simulations \cite{Geyer91} for our three polymer models.
Both canonical and microcanonical ensembles have been simulated. A
good sampling requires the knowledge of the global minimum for each
polymer. These have been found using the basin-hopping scheme \cite{Wales97,Li87}.
We confirm in this way that the geometrical structure of the native
state in the DHTP coincides with that of the HMP global minimum. It is
interesting to point out that the search of the DHTP protein native structure
is a factor of 10 faster than that of the HMP and RHTP global minima,
which is a direct consequence of their very different energy landscapes.%
\begin{figure}[b]
\psfrag{A}[tc][tc]{\large $(a)$}
\psfrag{B}[tc][tc]{\large $(b)$}
\psfrag{C}[tc][tc]{\large $(c)$}
\psfrag{cv}[bc][bc]{$C_v(T)/N k_B$}
\psfrag{4.0}[tc][tc]{$4.0$}
\psfrag{3.5}[tc][tc]{$3.5$}
\psfrag{3.0}[tc][tc]{$3.0$}
\psfrag{3.5}[tc][tc]{$3.5$}
\psfrag{3.0}[tc][tc]{$3.0$}
\psfrag{2.5}[tc][tc]{$2.5$}
\psfrag{8.0}[tc][tc]{$8.0$}
\psfrag{6.0}[tc][tc]{$6.0$}
\psfrag{4.0}[tc][tc]{$4.0$}
\psfrag{2.0}[tc][tc]{$2.0$}
\psfrag{5}[tc][tc]{$5$}
\psfrag{10}[tc][tc]{$10$}
\psfrag{15}[tc][tc]{$15$}
\psfrag{20}[tc][tc]{$20$}
\psfrag{25}[tc][tc]{$25$}
\psfrag{kT}[tc][tc]{$k_BT$}
\includegraphics[width=8.25cm]{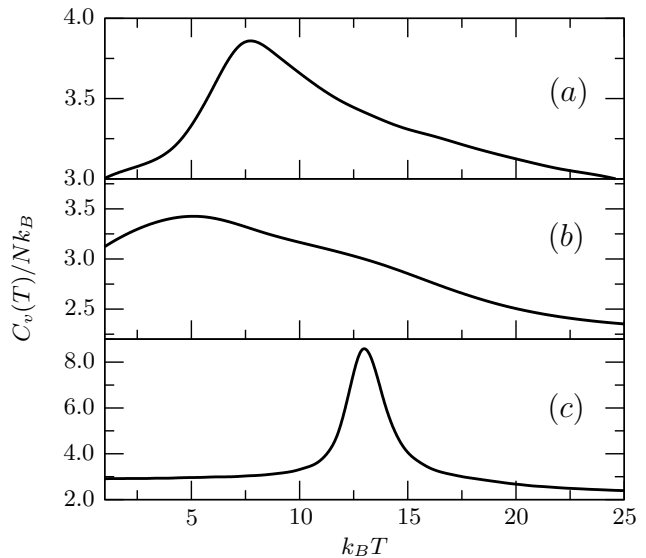}

\caption{\label{f1}Canonical heat capacities for the polymer models HMP
(a), RHTP (b) and DHTP (c).}

\end{figure}

Starting out from each of these global minima we perform 50 replicas.
For the canonical ensemble these replicas expand the temperature range
$0<k_{B}T\leq25$ ($k_{B}$ being the Boltzmann constant), while the
microcanonical sampling is performed in the energy interval $-1560\leq E\leq875$.
In order to equilibrate our systems at each sampling point, $1\times10^{8}$
Monte Carlo steps were run. Statistics were then collected for the
same number of additional steps. Results for the canonical heat capacity
$C_{v}(T)=\frac{3}{2}Nk_{B}+(\left\langle U^{2}\right\rangle -\left\langle U\right\rangle ^{2})/k_{B}T^{2}$,
where $U$ is the potential energy, are presented in Fig. \ref{f1}. 

The three polymer models show a single peak in the heat capacity that,
using the same energy scale, is narrower for the DHTP model.
Clearly, from these results we cannot extract any definite conclusion
about the nature of the transition associated with the heat capacity
peaks.

The situation changes dramatically when we move to the microcanonical
ensemble.%
\begin{figure}[t]
\psfrag{A}[tc][tc]{\large $(a)$}
\psfrag{B}[tc][tc]{\large $(b)$}
\psfrag{C}[tc][tc]{\large $(c)$}
\psfrag{Total Energy}[tc][tc]{$E$}
\psfrag{4}[tc][tc]{$4$}
\psfrag{6}[tc][tc]{$6$}
\psfrag{8}[tc][tc]{$8$}
\psfrag{-1200}[tc][tc]{$-1200$}
\psfrag{-1100}[tc][tc]{$-1100$}
\psfrag{-1000}[tc][tc]{$-1000$}
\psfrag{-900}[tc][tc]{$-900$}
\psfrag{-800}[tc][tc]{$-800$}
\psfrag{0}[tc][tc]{$0$}
\psfrag{3}[tc][tc]{$3$}
\psfrag{6}[tc][tc]{$6$}
\psfrag{9}[tc][tc]{$9$}
\psfrag{-600}[tc][tc]{$-600$}
\psfrag{-400}[tc][tc]{$-400$}
\psfrag{-200}[tc][tc]{$-200$}
\psfrag{0}[tc][tc]{$0$}
\psfrag{16}[tc][tc]{$16$}
\psfrag{14}[tc][tc]{$14$}
\psfrag{12}[tc][tc]{$12$}
\psfrag{10}[tc][tc]{$10$}
\psfrag{-450}[tc][tc]{$-450$}
\psfrag{-300}[tc][tc]{$-300$}
\psfrag{-150}[tc][tc]{$-150$}
\psfrag{0}[tc][tc]{$0$}
\psfrag{150}[tc][tc]{$150$}
\psfrag{300}[tc][tc]{$300$}
\psfrag{450}[tc][tc]{$450$}
\psfrag{kT}[tc][tc]{$k_BT$}
\psfrag{kT}[tc][tc]{$k_BT$}
\psfrag{kT}[tc][tc]{$k_BT$}
\includegraphics[width=8.25cm]{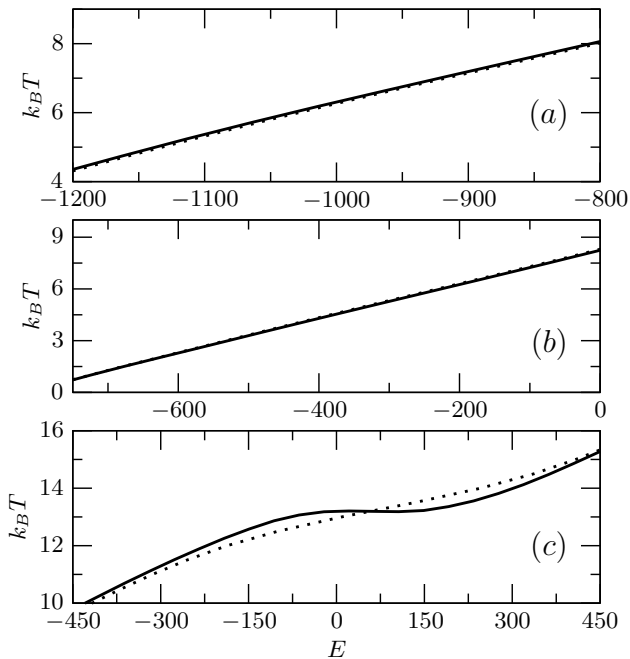}

\caption{\label{f2}Caloric curves for the polymer models HMP (a), RHTP (b), and DHTP (c).
Results are for the canonical (dotted line) and
microcanonical (full line) ensembles.}

\end{figure}
 Fig. \ref{f2} presents energy-temperature plots (caloric curves)
for both ensembles and three polymer models. While the data for both the
HMP and RHTP models show the same behavior in both ensembles, the S-bend
microcanonical feature that characterizes first-order-like phase transitions
is observed in the DHTP model. The origin of this feature
can be understood in terms of the two-phase picture behind the folding
transition. Since the native state corresponds to a very deep potential
well, the low temperature behavior will follow closely that of a multidimensional
harmonic oscillator, with a density of states that grows with energy
as a power law. Therefore, entropy is a concave function of energy.
At energies where the unfolded state starts to participate, large
high potential energy regions of the DHTP configuration space
are sampled; thus the density of states grows significantly and a
convex intruder appears in the above function. In other words, the
system becomes colder upon absorbing energy and the microcanonical
heat capacity (not shown here) becomes negative in the phase coexistence
region. The same behavior is not possible in the $\theta$-like transition
for for both HMP and RHTP models because, due to the rugged nature of the potential
energy landscape, the energy regions just above and below the transition 
zone are very much alike.%
\begin{figure}[t]
\psfrag{A}[tc][tc]{\large $(a)$}
\psfrag{B}[tc][tc]{\large $(b)$}
\psfrag{C}[tc][tc]{\large $(c)$}
\psfrag{P\(E\)}[bc][bc]{$P(E)\times 10^{3}$}
\psfrag{P\(E\)}[bc][bc]{$P(E)\times 10^{3}$}
\psfrag{P\(E\)}[bc][bc]{$P(E)\times 10^{3}$}
\psfrag{Energy}[tc][tc]{$E$}
\psfrag{0}[tc][tc]{$0$}
\psfrag{2}[tc][tc]{$2$}
\psfrag{4}[tc][tc]{$4$}
\psfrag{6}[tc][tc]{$6$}
\psfrag{8}[tc][tc]{$8$}
\psfrag{-1500}[tc][tc]{$-1500$}
\psfrag{-1350}[tc][tc]{$-1350$}
\psfrag{-1200}[tc][tc]{$-1200$}
\psfrag{-1050}[tc][tc]{$-1050$}
\psfrag{-900}[tc][tc]{$-900$}
\psfrag{0}[tc][tc]{$0$}
\psfrag{8}[tc][tc]{$8$}
\psfrag{-700}[tc][tc]{$-700$}
\psfrag{-600}[tc][tc]{$-600$}
\psfrag{-500}[tc][tc]{$-500$}
\psfrag{-400}[tc][tc]{$-400$}
\psfrag{0}[tc][tc]{$0$}
\psfrag{1}[tc][tc]{$1$}
\psfrag{2}[tc][tc]{$2$}
\psfrag{3}[tc][tc]{$3$}
\psfrag{-1000}[tc][tc]{$-1000$}
\psfrag{-800}[tc][tc]{$-800$}
\psfrag{-600}[tc][tc]{$-600$}
\psfrag{-400}[tc][tc]{$-400$}
\psfrag{-200}[tc][tc]{$-200$}
\psfrag{0}[tc][tc]{$0$}
\psfrag{6.3}[tc][tc]{$6.3$}
\psfrag{7.8}[tc][tc]{$7.8$}
\psfrag{9.3}[tc][tc]{$9.3$}
\psfrag{4.0}[tc][tc]{$4.0$}
\psfrag{5.0}[tc][tc]{$5.0$}
\psfrag{6.0}[tc][tc]{$6.0$}
\psfrag{13.0}[tc][tc]{$13.0$}
\psfrag{13.5}[tc][tc]{$13.5$}
\psfrag{14.0}[tc][tc]{$14.0$}
\includegraphics[width=8.25cm]{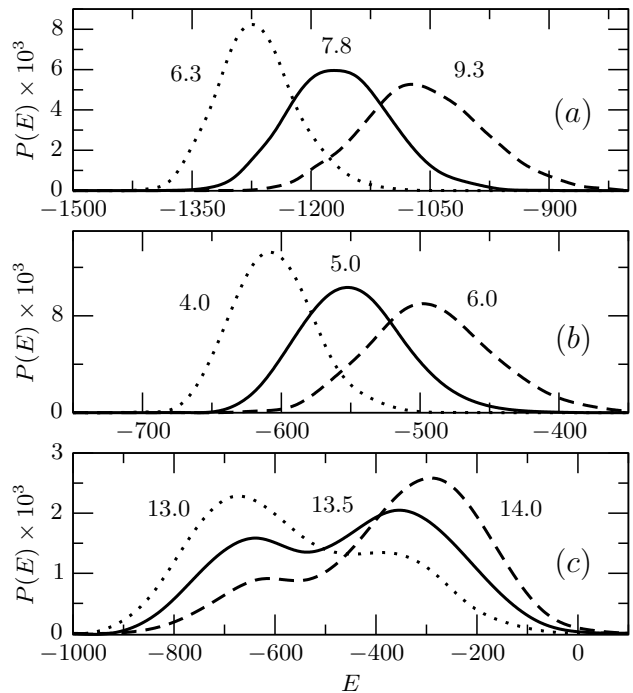}

\caption{\label{f3}Energy distribution functions. HMP (a), RHTP (b) and
DHTP (c). Numbers within the panels are the $k_{B}T$ values.}

\end{figure}

We have mentioned in the introductory paragraphs that the microcanonical
S-bend feature is directly connected to the bimodality of the energy
distribution function at the transition temperature. In Fig. \ref{f3}
we present this function for the three models
at three temperatures close to the corresponding
transition temperature. The totally different performance of these
models is clearly observed;
the two phase coexistence is obvious in the DHTP model. As is
well known, the energy difference between the two maxima in the bimodal
energy distribution provides the transition latent heat ($\Delta Q=2.8\times10^{2}$).
The energy distribution functions for both HMP and RHTP show also an interesting
behavior, namely, not only the average energy increases monotonically
with temperature but also the energy fluctuation does so. Of course,
DHTP shows a maximum of this fluctuation at the transition
temperature, as expected from its first-order-like nature. The internal
energy fluctuation is nothing but $[C_{v}(T)-\frac{3}{2}Nk_{B}]k_{B}T^{2}$;
therefore, when the heat capacity is properly shifted and multiplied
by $T^{2}$ the peaks observed in Fig. \ref{f1} for both HMP and RHTP
do indeed disappear, but the one for DHTP
does not. Note that the same behavior is observed respectively for
bad and good folders in the fluctuation of the energy
landscape curvature defined by
Mazzoni and Casetti \cite{Casetti06}. This measure is directly related
with the internal energy and the behavior found for the fluctuations
of the latter might as well be the origin of the similar behavior
observed in the fluctuation of this curvature.%
\begin{figure}[t]
\psfrag{dT/dE}[bc][bc]{$(k_BdT/dE)\times 10^{3}$}
\psfrag{x}[tc][tc]{$x$}
\psfrag{-4}[tc][tc]{$-4$}
\psfrag{-2}[tc][tc]{$-2$}
\psfrag{0}[tc][tc]{$0$}
\psfrag{2}[tc][tc]{$2$}
\psfrag{4}[tc][tc]{$4$}
\psfrag{6}[tc][tc]{$6$}
\psfrag{8}[tc][tc]{$8$}
\psfrag{0}[tc][tc]{$0$}
\psfrag{0.2}[tc][tc]{$0.2$}
\psfrag{0.4}[tc][tc]{$0.4$}
\psfrag{0.6}[tc][tc]{$0.6$}
\psfrag{0.8}[tc][tc]{$0.8$}
\psfrag{1}[tc][tc]{$1$}
\includegraphics[width=8.25cm]{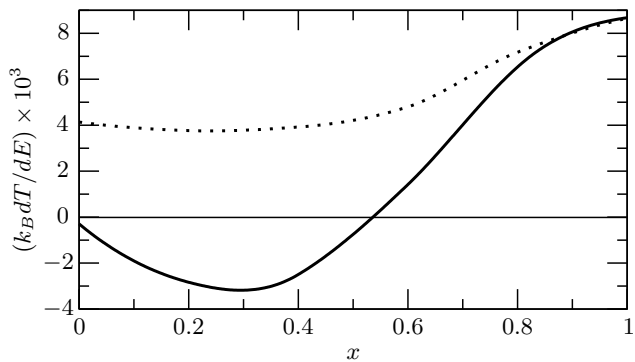}

\caption{\label{f4}Minimum values of the canonical (dotted line) and microcanonical (full line) caloric curve derivatives
as a function of $x$ (see text for details).}

\end{figure}

Finally, an interesting issue is concerned with the following question:
Once we have set the PES model parameters for a good folder, how much
can we change them towards those for the bad folder without essentially
altering the quality of the model as a good folder? In order to answer
this question we have repeated the previous microcanonical analysis
for a set of models with parameters chosen as $\{p\}_{x}=\{p\}_{{\rm g}}-(\{p\}_{{\rm g}}-\{p\}_{{\rm b}})x$
with $0\leq x\leq1$, where $\{p\}_{{\rm g}}$ and $\{p\}_{{\rm b}}$
are respectively the parameter sets for DHTP and
HMP models. Fig. \ref{f4} presents the minimum values
of $dT/dE$ from both the canonical and microcanical caloric curves.
Notice that the transition
from good ($dT/dE<0$) to bad ($dT/dE>0$) folder occurs at $x\sim0.54$,
and that the good folder behavior is markedly better for $x\sim0.3$.
The other properties characterizing good folders (unique native state and
funneled energy landscape) were checked to be linked to
the negative $dT/dE$ interval.
Therefore, even strong perturbations of the  parameters for a good folder
do not change the essential behavior of the folding transition.

Concluding, we have shown that the microcanonical analysis is a powerful
tool to characterize the protein folding transition and to neatly
distinguish between good and bad folders. We have applied this scheme
to three toy models representing these two types of polymers and have
shown that once we have a good folder, its properties are quite robust
against perturbations in the interaction parameters. Possible connections
between the features that come out from our analysis and the behavior
of measures such as the curvature of the energy landscape \cite{Casetti06}
have been discussed.

This work was supported by `Ministerio de Educaci\'on y Ciencia (Spain)'
and `FEDER fund (EU)' under contract No. FIS2005-02886. We thank Dr. F. Calvo
and Dr. D.J. Wales for their valuable comments.

\end{document}